\documentclass[sigconf]{acmart}

\AtBeginDocument{%
  }

\copyrightyear{2025}
\acmYear{2025}
\setcopyright{cc}
\setcctype{by}
\acmDOI{10.1145/3746059.3747662}
\acmConference[UIST '25]{The 38th Annual ACM Symposium on User Interface Software and Technology}{September 28-October 1, 2025}{Busan, Republic of Korea}
\acmBooktitle{The 38th Annual ACM Symposium on User Interface Software and Technology (UIST '25), September 28-October 1, 2025, Busan, Republic of Korea}
\acmISBN{979-8-4007-2037-6/2025/09}

\begin{document}

\newcommand{\Sys}{\textsc{Capybara}}
\newcommand{\add}[1]{\textcolor{black}{#1}}

\title{Empowering Children to Create AI-Enabled Augmented Reality Experiences}


\author{Lei Zhang}
\affiliation{%
 \institution{New Jersey Institute of Technology \newline \& Princeton University}
 \city{Newark}
 \state{NJ}
 \country{USA}}
 \email{lei.zhang@njit.edu}

\author{Shuyao Zhou}
\affiliation{%
 \institution{Princeton University}
 \city{Princeton}
 \state{NJ}
 \country{USA}}
 \email{sz8740@princeton.edu}

\author{Amna Liaqat}
\affiliation{%
 \institution{Princeton University}
 \city{Princeton}
 \state{NJ}
 \country{USA}}
 \email{al0910@princeton.edu}

\author{Tinney Mak}
\affiliation{%
 \institution{Princeton University}
 \city{Princeton}
 \state{NJ}
 \country{USA}}
 \email{tm0261@princeton.edu}

 \author{Brian Berengard}
\affiliation{%
 \institution{The Clubhouse Network}
 \city{Buenos Aires}
 \country{Argentina}}
 \email{berengard.brian@gmail.com}

 \author{Emily Qian}
\affiliation{%
 \institution{Princeton University}
 \city{Princeton}
 \state{NJ}
 \country{USA}}
 \email{eqian27@princeton.edu}

\author{Andrés Monroy-Hernández}
\affiliation{%
 \institution{Princeton University}
 \city{Princeton}
 \state{NJ}
 \country{USA}}
 \email{andresmh@cs.princeton.edu}

\renewcommand{\shortauthors}{Lei Zhang et al.}

\begin{abstract}
Despite their potential to enhance children's learning experiences, AI-enabled AR technologies are predominantly used in ways that position children as consumers rather than creators.
We introduce Capybara, an AR-based and AI-powered visual programming environment that empowers children to create, customize, and program 3D characters overlaid onto the physical world. 
Capybara enables children to create virtual characters and accessories using text-to-3D generative AI models, and to animate these characters through auto-rigging and body tracking. In addition, our system employs vision-based AI models to recognize physical objects, allowing children to program interactive behaviors between virtual characters and their physical surroundings. We demonstrate the expressiveness of Capybara through a set of novel AR experiences. We conducted user studies with 20 children in the United States and Argentina. Our findings suggest that Capybara can empower children to harness AI in authoring personalized and engaging AR experiences that seamlessly bridge the virtual and physical worlds.
\end{abstract}


\begin{teaserfigure}
  \includegraphics[width=\textwidth]{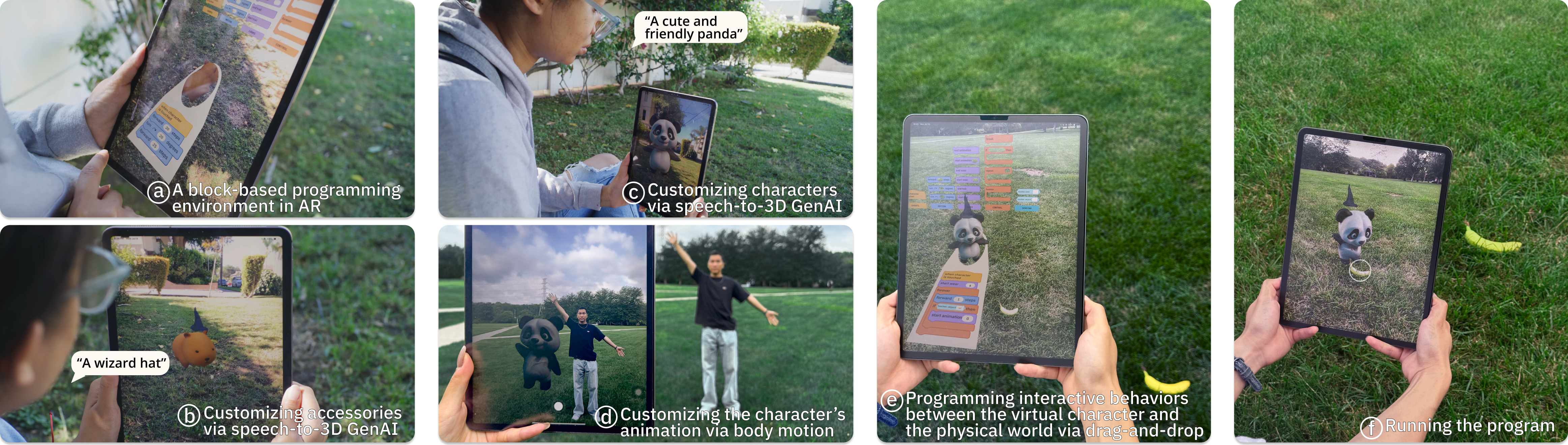}
  \vspace{-1.5pc}
  \caption{System Overview of Capybara. (a) Capyabara is a block-based programming environment in mobile Augmented Reality (AR) that empowers children to program AI-enabled AR experiences. (b \& c) Users can customize virtual 3D characters and accessories via the speech-to-3D GenAI feature. (d) The system auto-rigs the character and lets users control the character's animation via body motion. (e) Via drag-and-drop programming, users can define interactive behaviors between virtual characters and their physical surroundings using object detection. 
  \add{In this example, a cute panda character wearing a wizard hat waves their arms to keep balance when stepping on a physical banana.} (f) Combining the above functionalities, Capybara enables novel AR experiences that bridge the virtual and physical worlds.}
  \label{fig:teaser}
\end{teaserfigure}


\maketitle

\section{Introduction}
The advancement of Artificial Intelligence (AI) has enhanced Augmented Reality (AR) experiences, making them more intelligent, engaging, and personalized~\cite{suzuki2024everyday}.
Furthermore, researchers have integrated AI capabilities into AR to enrich children's educational and social experiences.
For example, AI-powered AR systems can transform static content from physical textbooks into interactive 3D diagrams, supporting more immersive and effective learning~\cite{gunturu2024augmented}.
Many interactive social applications, such as AR filters~\cite{dagan2022project}, leverage facial and body recognition to create more playful, engaging, and memorable social experiences for children.
Such experiences are made possible by AI's increasing capabilities to understand the physical environment (e.g., through computer vision) or to synthesize novel content (e.g., through generative AI).

Despite the recent technological advances, digital tools designed for children predominantly emphasize content consumption over creation, i.e., they focus on ``reading'' rather than ``writing.'' A notable exception is authoring tools for children such as Scratch~\cite{resnick2009scratch}, which are developed based on Constructionism, a pedagogy that emphasizes learning by building ~\cite{resnick1996pianos}. Constructionist tools aim to empower children to become creators of digital experiences for self-expression and creative exploration. 
However, to our knowledge, there are no Constructionist tools for children to create increasingly complex and expressive AR experiences enhanced by AI capabilities nowadays.
This is particularly challenging because creating such experiences typically require expertise from 3D modeling and animation, to programming and deploying AI models. 
For instance, creating 3D characters and their animations for storytelling often demands advanced skills from professional 3D modelers and animators. 
Programming interactions between the physical and virtual worlds requires not only 3D programming knowledge, but also an understanding of how to incorporate AI models effectively. 
In this research, we aim to empower children to be creators of AI-enabled AR experiences by equipping them with a novel authoring tool.

This paper introduces \Sys{}, an authoring tool for children to create AI-enabled AR experiences. Our system offers a block-based programming environment in AR that allows children to construct visual programs by dragging and dropping code blocks. 
\Sys{} raises the ceiling of expressive power by introducing three novel functionalities: 1) 3D character customization, 2) animation customization through puppeteering, and 3) programming interactions across virtual and physical worlds. Specifically, our system enables children to generate 3D characters using natural language descriptions, powered by a GenAI model that supports text-to-3D synthesis. 
Children can also generate accessories for their character using the same technique, as well as programmatically create and destroy these accessories. 
In addition, our system can automatically rig these generated 3D characters, enabling children to control their animation via body motion and programmatically replay those custom animations. These features offer a seamless connection between the virtual character and the user's physical body. 
Finally, our system enables children to utilize off-the-shelf AI functionalities, such as object recognition, to program interactions between the virtual character and the physical environment.

Combining these techniques into the block-based programming environment, \Sys{} empowers children to create expressive, AI-enabled experiences. 
To demonstrate the system's expressiveness, we present a range of example experiences created using \Sys{}. 
To evaluate the usability of our system, we conducted user studies with 20 children aged 7 to 16 across the United States and Argentina. Our findings suggest that \Sys{} enabled our participants to create AI-enabled AR experiences. We found that \Sys{} enabled creative usage such as \add{generating 3D characters for both realistic and fantastical experiences, animating the character for storytelling, and building interactions between the virtual and physical worlds. Participants found the app to be unique and fun. They felt that it could facilitate creative expression through AI-enabled customization, enable programming interactions with the physical world, and engage them with concepts in Computational Thinking (CT) and AI literacy.} However, integrating AI into the authoring process could introduce alignment issues \add{(e.g., mismatches between AI results and children's intent)} and distract children from their creation goals. These findings offer insights into designing AI-enabled AR authoring tools for children to foster creativity and self-expression.  

In summary, our paper makes the following contribution: 1) \Sys{}, an authoring tool for children to create AI-enabled AR experiences, which seamlessly combines AI techniques with children's content creation process; 2) a set of example experiences demonstrating the expressiveness of the tool; and 3) empirical insights gained from our user studies with children, highlighting its benefits and challenges of using \Sys{}.

\section{Related Work}

We draw from prior literature in AR/VR authoring tools, visual programming languages, and Children-AI Interaction.

\subsection{Authoring Tools for Immersive Experiences}
Authoring AR/VR experiences has been known to have a high barrier of entry for end-users such as people with few technical skills~\cite{ashtari2020creating}.
In recent years, many novel AR/VR prototyping tools have been proposed to lower the ``floor'' of creating immersive experiences~\cite{lee2004immersive, nebeling2018protoar, nebeling2019360proto, zhang2020flowmatic, suzuki2020realitysketch, leiva2021rapido, wang2021gesturar, chen2021scenear, hedlund2023blocklyvr, frau2023xrspotlight, ye2024prointerar}.
A key challenge is to enable the creation of interactive behaviors, which typically require extensive programming knowledge.
Visual programming languages have shown to be an effective approach for people with less technical skills to define interactive behavior.
For example, FlowMatic~\cite{zhang2020flowmatic} enables users to define interactive behaviors using dataflow diagrams in VR.
ProInterAR allows users to create interactive AR scenes by constructing block-based visual programs with the combination of a tablet and AR-HMD~\cite{ye2024prointerar}.
More recently, with the rise of generative AI, there have been attempts to create immersive experiences with generative models~\cite{liao2022realitytalk, monteiro2023teachable, de2024llmr, giunchi2024dreamcodevr, zhang2024vrcopilot, aghel2024people, jennings2024s}.
Specifically, natural language descriptions have been explored to enable end-users to quickly prototype interactive AR/VR scenes (e.g., ~\cite{jennings2024s}).
Teachable Reality can identify user-defined tangible and gestural interactions using an on-demand computer vision model and enable the creation of AR interactions~\cite{monteiro2023teachable}.
However, the above tools were targeted for adult designers or novice programmers rather than children. 

As AR started to gain momentum in educational settings~\cite{billinghurst2002augmented}, recent research has started to recognize the importance of enabling children, e.g., K-12 students, to author or customize their own AR experiences~\cite{silva2023development, zhang2025following}.
A few authoring tools were introduced to empower children to be creators of AR experiences~\cite{radu2009augmented, annett2015moveablemaker, singh2021story, lunding2022exposar, chen2024oh}. For example, AR Scratch allows children to use the Scratch platform to create AR experiences~\cite{radu2009augmented}. MovableMaker enables children to design and customize movable effects via animated tooltips, automatic instruction generation, and constraint-based rendering~\cite{annett2015moveablemaker}. ExposAR allows children to collaboratively author AR experiences via image and plane recognition~\cite{lunding2022exposar}.
A key limitation of these existing authoring tools for children is their low ceiling of expressiveness for creating complex and expressive AR experiences enhanced by AI capabilities.
As AI-enabled AR experiences are becoming increasingly prevalent~\cite{suzuki2023xr, hirzle2023xr}, our work contributes to existing literature by empowering children to use AI techniques to author their AR experiences.
Specifically, \Sys{} raises the ceiling of expressiveness by enabling children to customize the AR characters via GenAI, to customize the animation of the characters via puppeteering, and to program the interactions between the virtual character and the physical world.

\subsection{Visual Programming Languages}
While there are multiple visual programming languages designed for non-programmers, block-based programming languages are one of the most popular platforms for children in computing education.
These languages use visual code blocks that represent code concepts, making it easier, especially for beginners, to learn programming by snapping blocks together instead of writing syntax~\cite{weintrop2019block}.
Over the decades, a few block-based programming environments have been introduced for teaching computational thinking in children (e.g., ~\cite{conway2000alice, kelleher2007storytelling, resnick2009scratch, blockly}).
For example, Scratch, as one of the most influential block-based programming environments, has empowered children to express themselves~\cite{resnick2009scratch}.

More recently, several extensions to block-based programming languages were proposed to improve teaching computational thinking among children (e.g., ~\cite{dietz2023visual, chen2024chatscratch, chen2024coremix}). For example, Visual StoryCoder proposes a voice-based agent to deliver instructions to children while they are programming using code blocks without an instructor~\cite{dietz2023visual}. ChatScratch built on this work by utilizing genAI, i.e., a stable diffusion model for image generation, to help children in the creative coding process~\cite{chen2024chatscratch}. However, these tools primarily focus on authoring digital experiences on 2D screens, such as 2D assets, characters, and animation. 

Our work contributes to this line of research by exploring the design of a block-based programming environment in AR for children. Specifically, we explore AR's unique spatial aspect by enabling children to manipulate code blocks in AR, customize characters via text-to-3D and body tracking, and program the interactions between virtual characters and the physical world.

\subsection{Children-AI Interaction}
\add{AI is transforming how children learn and create~\cite{davis2016empirically, long2020ai, druga2022family, druga2021children, touretzky2019envisioning}.
Long and Magerko highlighted the need for fostering AI literacy in audiences without technical background, and defined AI literacy by proposing a set of competencies that enables children to critically evaluate AI technologies, and communicate and collaborate effectively with AI~\cite{long2020ai}.
AI literacy is also considered a key component of computational thinking~\cite{touretzky2019envisioning}, which can cultivate more general problem-solving skills.
With the recent rise of GenAI models, Kosoy and colleagues found that children generally hold a positive outlook on GenAI, expressing enthusiasm about its potential to enhance their everyday lives~\cite{kosoy2024childrensmentalmodelsgenerative}.}
A growing body of research \add{has explored} how GenAI can foster creativity in children ~\cite{zhang2022storybuddy, Zhang_2024, newman2024want}. For example, Mathemyths leverages Large Language Models (LLMs) to create a child–AI co-creative storytelling experience that integrates in-context explanations of mathematical vocabulary~\cite{Zhang_2024}. Similarly, Newman and colleagues investigate how children aged 7 to 13 years interact with GenAI tools, specifically DALL·E and ChatGPT, examining how these interactions shape the creative process and the perceptions of AI among children~\cite{newman2024want}.
However, there is limited exploration of how to integrate GenAI capabilities to the content creation process for AR experiences, particularly in \add{the context} of fostering children creative expressions.
AR offers unique spatial and multimodal interaction affordances that align well with children’s creative practices~\cite{liaqat2023exploring}.
To address this gap, we integrate GenAI into AR to support co-creation experiences between children and AI.
We also use natural language and body movements as inputs \add{that support} dynamic and real-time interactions.

Meanwhile, computer vision models have become key enablers for AR experiences, allowing virtual content to interpret and interact with physical objects in the user’s environment ~\cite{chulpongsatorn2023augmented, gunturu2024augmented, naji2024augmented, rumbelow2024promise, kaviyaraj2025augmented}. 
\add{For instance,} Naji and colleagues incorporated YOLOv4 for real-time object detection, overlaying corresponding English words to support vocabulary learning among primary school students~\cite{naji2024augmented}. 
Blockplay.ai uses a neural network-based object recognition system to identify physical math manipulatives, examining how children aged 5–6 engage with the system to support early math learning~\cite{rumbelow2024promise}. 
Kaviyaraj and Uma introduced an image-generation pipeline to enhance object detection using deep learning, evaluating YOLOv7 within educational AR applications~\cite{kaviyaraj2025augmented}.
Their findings indicate improved test scores among fourth-grade students using the system.
However, these systems primarily rely on loading pre-made 3D models and inserting them into the scene. Our approach enables dynamic children–AI interaction with real-world objects by leveraging real-time object recognition from the live camera feed, allowing AR entities to respond meaningfully to physical objects that children choose to explore.
This design enables children to co-create with AI in ways that are grounded in their immediate physical contexts.

\section{The \Sys{} System}

\Sys{} is an iOS application designed for tablets and phones that enables users to program a digital character, by default a cute capybara, using drag-and-drop code blocks. We chose the capybara character because of its  ``cuteness'' and its popularity among young people~\cite{shteyngart2025}. Overall, \Sys{} empowers children to create AI-enabled Augmented Reality (AR) experiences by customizing characters and accessories using genAI, animating the character via body motion, and programming interactive behaviors between virtual and physical worlds using vision-based AI models.
We have three primary design goals for \Sys{}:
\begin{itemize}
    \item \textbf{D1 - Lowering the barrier of entry for programming AI-enabled AR behaviors.} A primary goal of \Sys{} is to lower the \textit{floor} for children to author AI-enabled AR experiences. As discussed in Sections 2.1 and 2.2, visual programming languages provide an intuitive and accessible medium for young learners to create digital content. Block-based environments such as Scratch~\cite{resnick2009scratch}, grounded in Constructionist principles of learning through making~\cite{resnick_pianos_1996}, have proven particularly effective. Building on this foundation, we aim not only to keep a low \textit{floor} of entry but also to raise the \textit{ceiling} by empowering children to create increasingly complex and expressive AR experiences enhanced by AI capabilities. \add{We designed \Sys{} to balance AI automation to support creative expression and preserving children's agency and opportunities for learning.}
    
    \item \textbf{D2 - Enabling higher customizability.} Prior work has shown that customizability can facilitate creative expressions among users~\cite{frich2019mapping, zhang2022auggie}. In the context of educational experiences for children, prior research has explored constructionist toolkits designed for children to support self-expression and creativity~\cite{davis2013education, dasgupta2016remixing}. This emphasis on customization also aligns with recent effort to enable static 2D character personalization using AI-generated images~\cite{leong2025paratrouper}. With \Sys{}, we aim to enable greater customizability of virtual 3D characters and their animations or motions, enabling children to create more memorable, expressive, and personalized AR experiences.
    
    \item \textbf{D3 - Facilitating interactions between the virtual and physical worlds.} As noted by Billinghurst~\cite{billinghurst2002augmented}, a key affordance of AR technology in education lies in its ability to support seamless integration between real and virtual environments. Recent studies with children have similarly highlighted their desire for richer interactions that could bridge the virtual and physical worlds~\cite{liaqat2024understanding}. In \Sys{}, we aim to facilitate such interactions both during the authoring process and in the resulting experiences. Specifically, during authoring, we aim to offer intuitive and tangible interaction metaphors that children are familiar with in the physical world. In the authored experiences, virtual characters can have meaningful interactions with physical objects, based on the system's understanding of the physical environment.
\end{itemize}

\begin{figure}[t!]
  \centering
  \includegraphics[width=\linewidth]{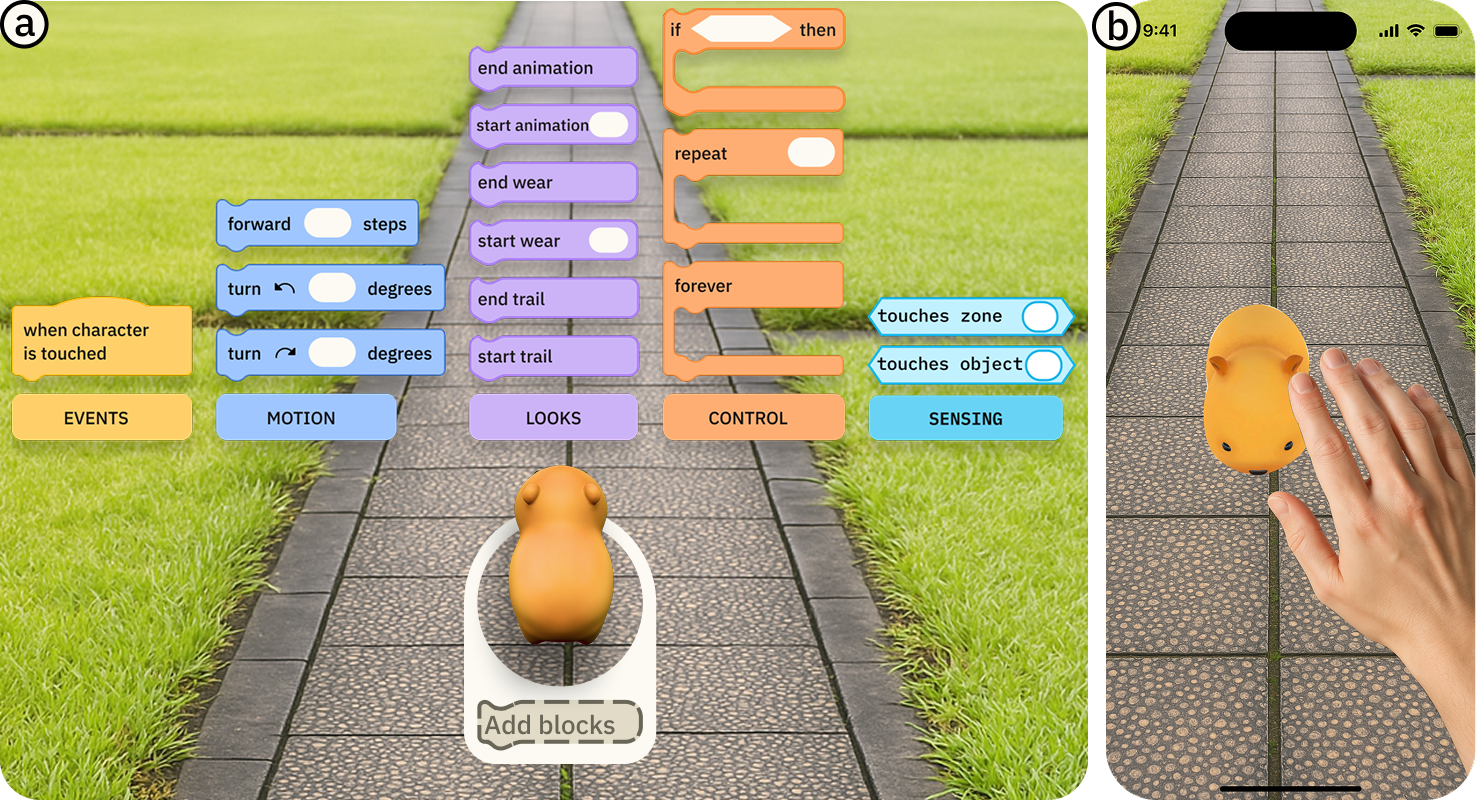}
  \vspace{-1.5pc}
  \caption{The Block-based Programming Environment of \Sys{}. (a) In Edit mode, \Sys{} visualizes five categories of code blocks and a canvas around the character for program construction in AR. (b) In Run mode, users execute the program by sticking out their hand to touch the character in AR, mimicking a patting action in the physical world.}
  \vspace{-0.5pc}
  \label{fig:block-based}
\end{figure}

\subsection{Block-based Programming Environments}
\Sys{} is based on a block-based programming environment in AR, as seen in Fig. \ref{fig:block-based}.
The code blocks feature an interlocking, puzzle-like design that visually guides users in constructing functional programs. 
These blocks are categorized into five distinct groups (see Fig.\ref{fig:block-based}.a), each with a unique color for easy identification:
\begin{enumerate}
    \renewcommand{\labelenumi}{(\theenumi)}
    \item \textbf{Event Blocks (Yellow)}: The Event block acts as a trigger that initiates the program’s execution. In \Sys{}, there is one event block: \texttt{When character is touched}. The user activates the program by sticking out their hand in front of the rear camera to ``touch'' the virtual character, mimicking the action of patting, as illustrated in Fig. \ref{fig:block-based}.b. 
    \item \textbf{Motion Blocks (Dark Blue)}: These blocks control the capybara’s movement, including translation and rotation. For example, the \texttt{forward} block moves the character forward by a specified number of steps, while the \texttt{turn} block rotates the character by a given number of degrees. Users can adjust the movement parameters using sliders for step count and rotation angle.
    \item \textbf{Looks Blocks (Purple)}: These blocks add visual elements to the character through trails, accessories, and animations. For instance, \texttt{start trail} creates a yellow paint trail as the capybara moves, allowing users to express 3D drawings in AR. Users can further personalize the character by applying accessories with the \texttt{start wear} block (see Section 3.2) and by customizing animations using the \texttt{start animation} block (see Section 3.3).
    \item \textbf{Control Blocks (Orange)}: These blocks introduce fundamental programming constructs such as loops and conditional logic, enabling more advanced behavior. The \texttt{repeat} block enables the code within to run as many times as indicated in the input box, while the code within the \texttt{forever} block will run repeatedly until the user manually stops it by tapping the character. The \texttt{if} block allows for conditional execution. These control blocks act as containers that execute the code blocks nested within them.
    \item \textbf{Sensing Blocks (Light Blue)}: These blocks enable the system to understand the physical world. The \texttt{touches object} block allows users to select a physical object and detect collision between the virtual character and that object (see Section 3.4.1). The \texttt{touches zone} block lets users define a region on a physical surface, enabling collision detection between the virtual character and the specific area (see Section 3.4.2).
\end{enumerate}
Upon launching, the app instructs users to place the character in the physical environment by finding a flat open area. 
Users then start coding by dragging and dropping code blocks from the code blocks inventory onto a canvas surrounding the character in AR.
The code blocks inventory hovers vertically above the canvas for easy access when the canvas is open.
Users can drag the code blocks away from the canvas to remove them.

\Sys{} distinguishes between on-screen interactions for programming and off-screen gestures for triggering actions. 
Users build programs through familiar touchscreen interactions, such as tapping to open the canvas and dragging code blocks. In contrast, program execution is initiated through playful gestures, such as mimicking a petting motion to trigger ``When character is touched.''
This clear separation between building and playing simplifies the programming process, while adding playfulness to the interaction with the character and the physical world.
By leveraging the affordances of the infinite AR play space and the block-based programming environment, \Sys{} aims to lower the entry barrier for children to begin creating their own AR experiences (\textbf{D1}).


\subsection{Customizing 3D Characters via GenAI}

We integrated Generative AI (GenAI) functionalities to allow children to create personalized AR experiences through customized \textit{characters} and \textit{accessories}. We designed it with the goals of stimulating users' creativity by enabling higher customizability (\textbf{D2}), while maintaining child safety standards.
The character customization process is initiated through the Settings menu, where users can select “Custom Character” to begin. Accessory customization, on the other hand, is triggered directly within the inventory interface using the \texttt{start wear} code block.

\begin{figure}[t!]
  \centering
  \includegraphics[width=\linewidth]{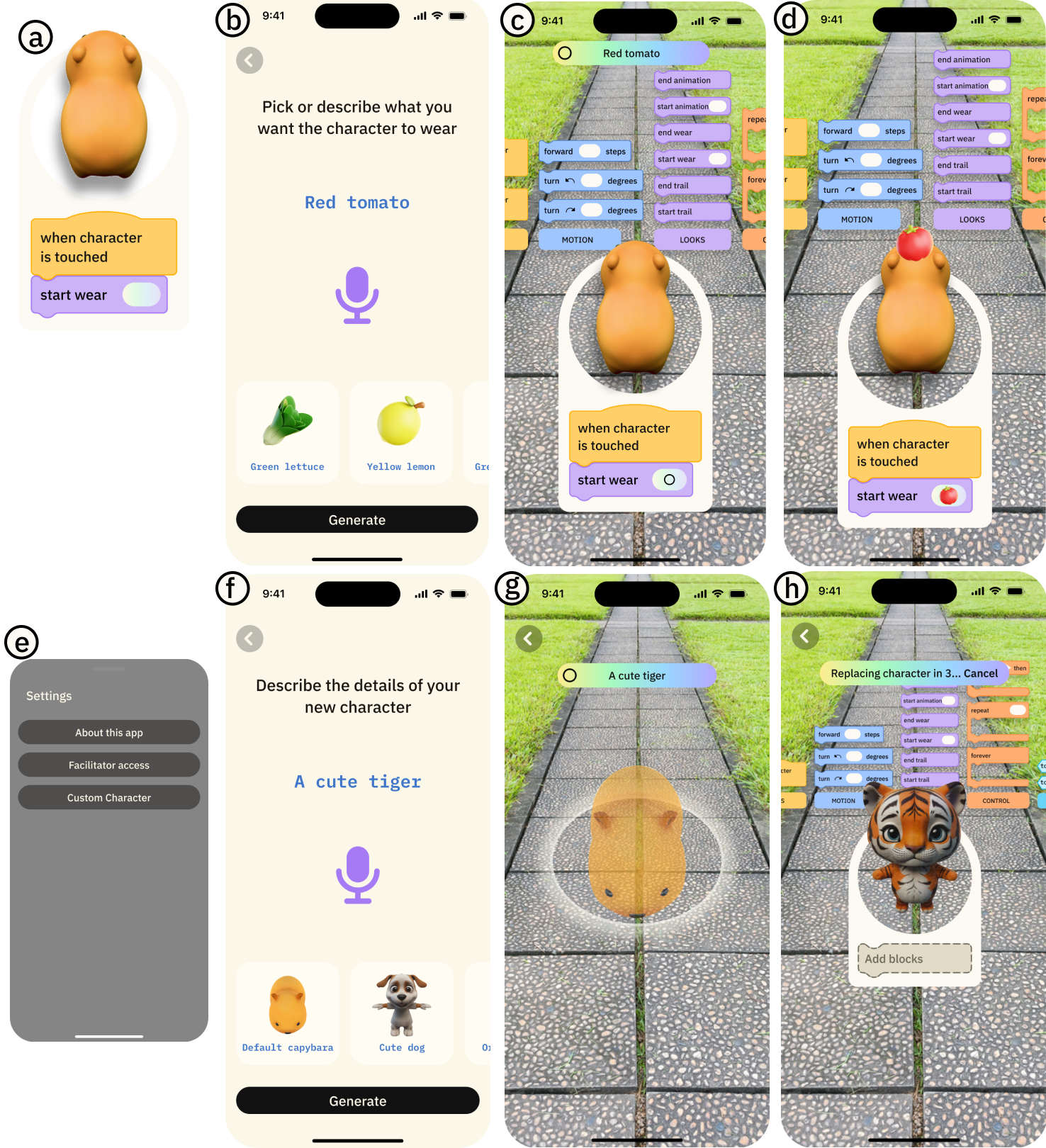}
  \vspace{-1.5pc}
  \caption{User interface for customizing accessories (a-d) and characters (e-h). (a) Users start by adding a code block named \texttt{start wear}. (b) The app jumps to a page for entering speech prompt or selecting from the library panel. (c) While generating the accessory, users can keep coding. (d) The accessory appears once generated. (e) Users start from the setting page to customize characters. (f) Users give a speech prompt. (g) The banner notification persists when waiting for GenAI completion. (h) The new character appears once generated.}
  \vspace{-.5pc}
  \label{fig:character_customization}
\end{figure}

Both character and accessory generation workflows follow a similar UI pattern, as shown in Fig. \ref{fig:character_customization}. Upon initiating the generation process, users see a dedicated creation page asking them to describe their desired character or accessory using speech.
For character customization, this page displays a panel of pre-defined characters, which includes the default capybara and predefined characters (past generations of characters are saved locally and added to the panel), to offer immediate options without API use. 
Similarly, the accessory generation page displays a set of pre-defined accessories. For original creations, users can press-and-hold the microphone icon to record speech input, which is then transcribed into text. The system uses the Google Gemini \add{1.5 Flash} LLM to review the prompt and ensure the appropriateness of the request for a young audience. If the initial prompt is flagged as inappropriate the system asks users to try again. Once the prompt passes the filter, the system calls an external text-to-3D-object generation service. The 3D object is displayed as a new character that the user can manipulate and program in the app's AR environment.

\subsection{Customizing Animation via Puppeteering}
While automating the rigging or skinning process of 3D models has been a popular research topic in the field of Computer Graphics (e.g., ~\cite{baran2007automatic, le2014robust, rignet}), it is unclear how these approaches can be seamlessly integrated into children's creative process of immersive experiences.
At the same time, research in the field of Human-Computer Interaction (HCI) has explored puppeteering \add{techniques} using controls such as human hand~\cite{hung2022puppeteer, jiang2023handavatar, hashimoto2024selfrionette} and body motion~\cite{yamane2010animating, seol2013creature}. However, the controlled 3D models were typically pre-rigged.
In \Sys{}, we refined and modified the open-sourced auto-rigging algorithm proposed by Baran and Popović~\cite{baran2007automatic}.
This is because the algorithm is computationally lightweight to be run on mobile devices and has been actively used as a baseline for recent algorithms that are more computationally heavy using neural networks (e.g., ~\cite{rignet}). 

In \Sys{}, we explore the creative process that integrates the complete pipeline from generating 3D characters (as shown in Section 3.2), to auto-rigging the character and controlling the character's animation via body motion.
Body motion is known to be an expressive way of creating personal and unique experiences in AR~\cite{saquib2019interactive}. We \add{designed} the puppeteering functionality \add{for children to control the character's animation}, with the goal of further enhancing customizability and facilitating \add{their} creative expressions (\textbf{D2}).
In addition, we chose body motion as a medium for puppeteering and let the virtual character mirror the user's motion to provide a seamless transition between the physical and the virtual (\textbf{D3}) during the authoring process.

\begin{figure}[t!]
  \centering
  \includegraphics[width=\linewidth]{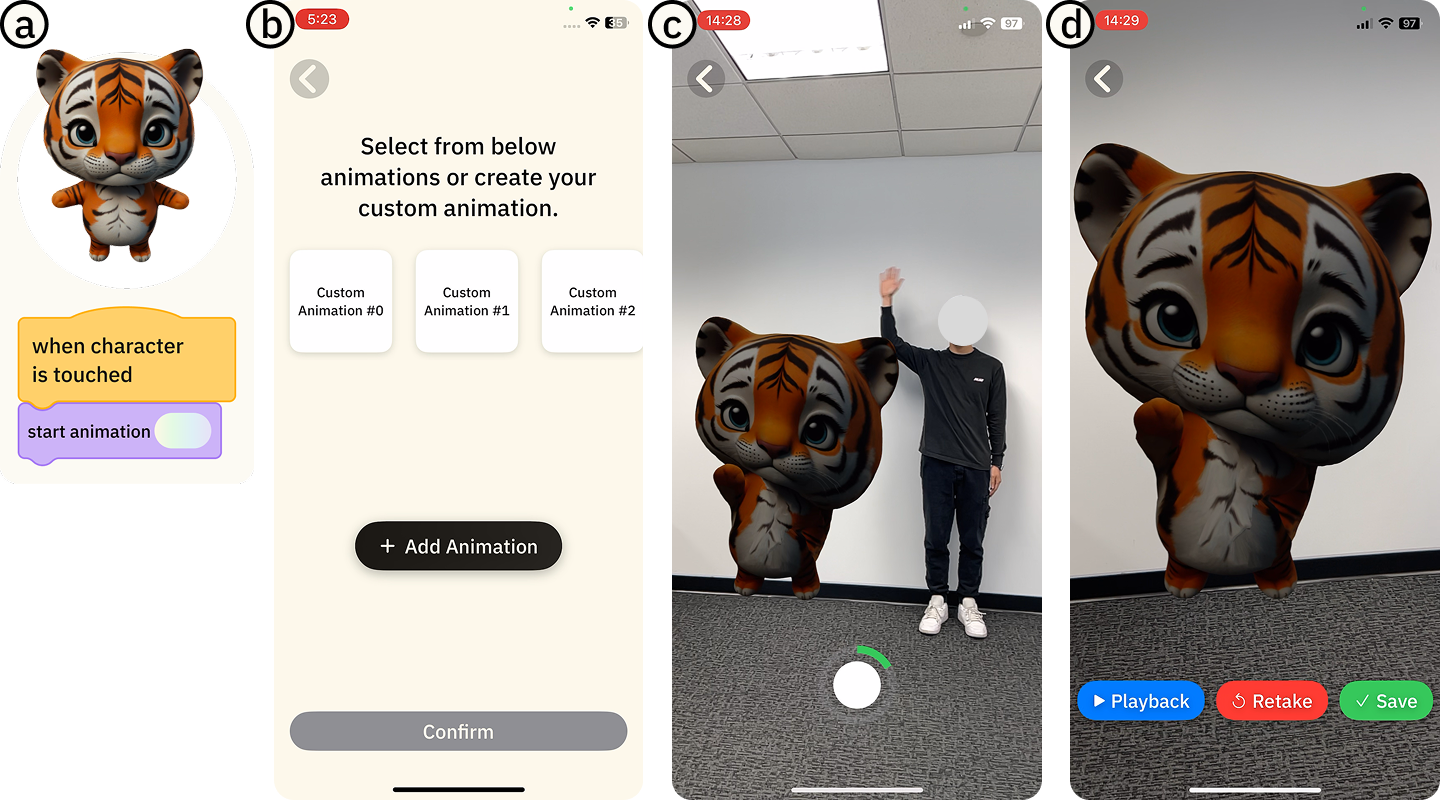}
  \vspace{-1.5pc}
  \caption{User interface for customizing animation. (a) Users start by adding a new code block named \texttt{start animation}. (b) The user is shown a page where they can select an existing animation from the top panel or add a new animation. (c) Users can hold the record button for recording an animation. (d) Users can playback immediately for review, retake, or save the animation for programmatic replay in the future.}
  \vspace{-0.5pc}
  \label{fig:animation_customization}
\end{figure}

In \Sys{}, the animation feature is integrated into the block-based programming environment by providing the code block named \texttt{start animation}. 
\add{Users can drag the codeblock and place it on the canvas, which then guides them to define the animation clip that can be programmatically replayed, as shown in Fig. \ref{fig:animation_customization}.}

\subsection{Bridging the Virtual and Physical World via Vision-based AI Models}
A key limitation of existing authoring tools for children is that the outcome experiences remain in the digital world. One of the design goals of \Sys{} is to facilitate children to create interactions between the virtual and the physical world (\textbf{D3}). We leverage off-the-shelf vision-based AI models to detect objects in the user's physical surroundings (Section 3.4.1). We also enable users to create self-defined zones to preserve users' agency in defining their own physical space and interaction (Section 3.4.2).

\subsubsection{Object Detection}
\begin{figure}[t!]
  \centering
  \includegraphics[width=\linewidth]{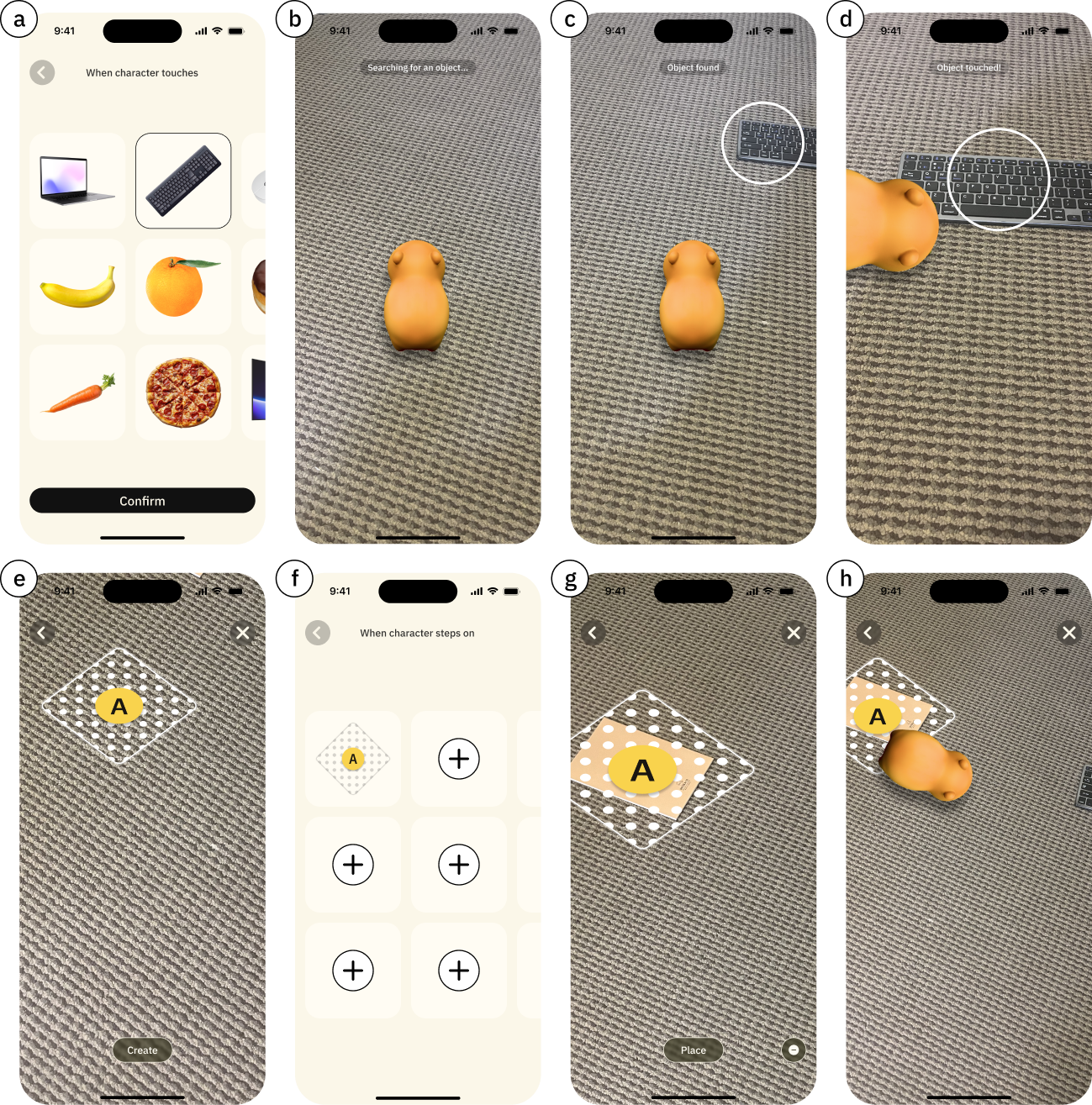}
  \vspace{-1.5pc}
  \caption{User interface for setting up object and zone detection. (a) Object selection. (b) Searching for the selected object in the view. (c) Object found in the view. (d) Character touches the object. (e) Zone creation. (f) Zone selection. (g) Zone placement. (h) Character steps on the zone.}
  \vspace{-.5pc}
  \label{fig:object_zone_combine}
\end{figure}

Users can program interactions with physical objects using the \texttt{touches object} block.
When users drag this block onto the canvas, they are directed to the page where they can select objects for detection (see Fig. \ref{fig:object_zone_combine}.a).
We adapted the YOLOv11s model \cite{Jocher_Ultralytics_YOLO_2023}, a deep neural network for real-time object detection. 
We curated a set of 24 recognizable objects, including food items, utensils, furniture, and common workplace tools, from the YOLOv11s detection classes. These objects were selected in order to support creative expressions of children's everyday experiences.

When the user runs the program that contains \texttt{touches object}, \Sys{} initiates real-time detection and displays a “Searching for object” message on the UI (see Fig.\ref{fig:object_zone_combine}.b). 
Once the object is found in the scene, the app overlays a flashing bounding box on the object and updates the UI to display “Object found” (see Fig.\ref{fig:object_zone_combine}.c). 
If a collision between the character and the selected object is detected, the UI displays “Object touched” (see Fig.\ref{fig:object_zone_combine}.d), and the subsequent code blocks are executed accordingly.

\subsubsection{Zone Detection}
To provide greater flexibility in programming interactions with the physical world, we also introduced the \texttt{touches zone} block as part of the Sensing Blocks.
Zones are user-defined regions on a physical surface that also support collision detection, enabling interactions with parts of the physical surroundings that are not covered by object detection.
Programming this block involves three views: zone creation, zone selection, and zone placement. 
When users drag this block onto the canvas, they are directed to the zone creation view, where a virtual rectangular zone is displayed on a detected surface. 
They can resize or manipulate the zone, and then finish creating the zone by tapping the \textit{Create} button (see Fig. \ref{fig:object_zone_combine}.e). 

After creating a zone, the user is directed to the zone selection view where they can see all the zones they have created, and select which zone to use for collision detection.
They can either choose from existing zones or add new ones by tapping the plus icon (see Fig. \ref{fig:object_zone_combine}.f). 
The system labels zones alphabetically (A to Z). 
After selecting a zone for collision detection, users enters the zone placement view, where they can adjust and confirm its location or size before tapping the \textit{Place} button (see Fig. \ref{fig:object_zone_combine}.g). 
The code block works by checking whether the character’s position falls within the boundaries of the selected zone. If so, the subsequent code blocks are executed (see Fig. \ref{fig:object_zone_combine}.h).

\begin{figure*}[t!]
  \centering
  \includegraphics[width=\linewidth]{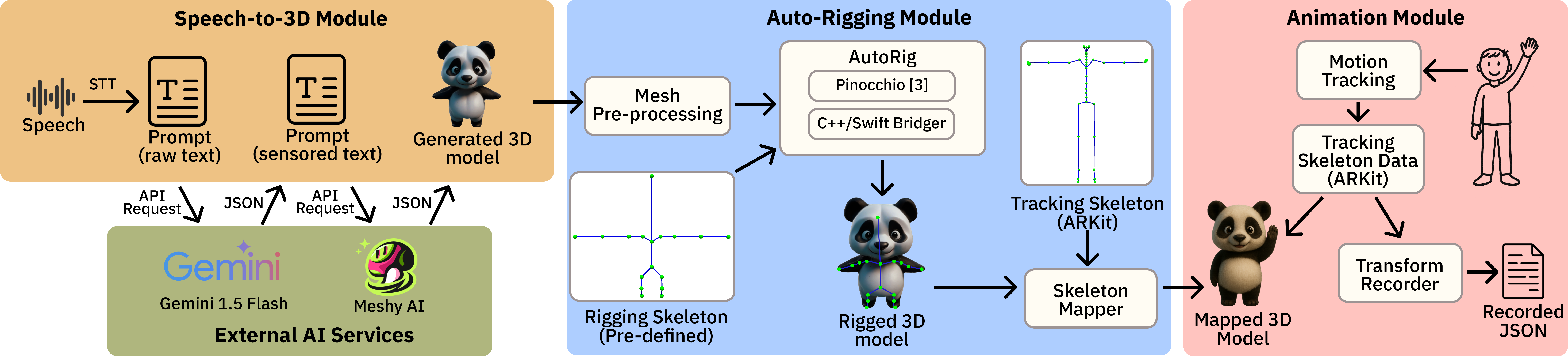}
  \vspace{-1pc}
  \caption{Character \& animation customization pipeline, which consists of three modules: speech-to-3D, auto-rigging, and animation. The speech-to-3D module takes a user's spoken prompt, filters it using an LLM, and generates a corresponding 3D model. The auto-rigging module ensures mesh connectivity in pre-processing, rigs the model using a predefined skeleton, and maps it to an ARKit-compatible skeleton. The animation module uses the camera feed to track human poses in real time, drives the 3D model with this skeleton data, and records the skeleton transformations as JSON for replay.}
  \vspace{-.5pc}
  \label{fig:pipeline}
\end{figure*}

\subsection{Implementation Details}

Overall, \Sys{} is an iOS app implemented using SwiftUI, ARKit, and RealityKit. In this section, we describe the implementation details, including customizing 3D characters via GenAI (Section 3.5.1), customizing animation via puppeteering (Section 3.5.2), and understanding and connecting with physical worlds (Section 3.5.3). A high-level architecture for character and animation customization is shown in Figure \ref{fig:pipeline}.

\subsubsection{Character Customization.}
\Sys{} enables users to customize 3D models such as characters and accessories via speech-to-3D functionalities, as shown in Figure \ref{fig:pipeline}.a. First, we implement the speech-to-text (STT) functionality via on-device speech recognition service provided by Apple\footnote{https://developer.apple.com/documentation/speech}. We use a Large-Language Model (LLM), i.e. Google Gemini 1.5 Flash, to moderate the user's prompt. The instruction we give to the LLM can be seen in the Appendix \add{A}. The moderation result is encoded into a JSON object that includes two fields: \texttt{inappropriateForChildren} and \texttt{reason}. The system will only proceed to generate the 3D model if the return value of \texttt{inappropriateForChildren} is \texttt{false}; otherwise, the system will ask the user to try a different prompt. Then the system sends the prompt through an API request to an external text-to-3D service, i.e. Meshy AI\footnote{https://meshy.ai/}. The returned JSON from the service includes a download link for the generated 3D models. The inference time for the text-to-3D service varies based on subscription level and network traffic, which typically takes less than five minutes in our studies. 

\subsubsection{Animation Customization.}
To enable users to customize the character's animation, our system employs an auto-rigging module and an animation module via puppeteering, as seen in Fig \ref{fig:pipeline}.
Specifically, our auto-rigging module is implemented based on the open-sourced algorithm named Pinocchio proposed by Baran and Popović~\cite{baran2007automatic}.
The module first conducts a pre-processing that checks the connectivity of the 3D mesh. 
\add{We then rig the generated 3D model using a pre-defined skeleton. Our auto-rigging algorithm assumes that the character is proportioned roughly like the pre-defined rigging skeleton (see the Limitations section for details). In edge cases where this assumption fails, such as generated meshes with extremely long or short limbs, or non-humanoid shapes, the rigging accuracy may degrade and users are currently expected to retry with a different model.
To enhance the robustness of the rigging process, we employed prompt engineering strategies to reduce the likelihood of generating unriggable models (see Appendix B).}
The on-device auto-rigging algorithm typically takes less than five minutes to run on an iPad Pro (M4 Chip, 2024).
Once the algorithm is completed, a Skeleton Mapper will replace the rigging skeleton with a tracking skeleton that is defined by ARKit.
This skeleton mapping is necessary for matching ARKit's motion capture requirements\footnote{https://developer.apple.com/documentation/arkit/validating-a-model-for-motion-capture}.
We chose to use a pre-defined rigging skeleton instead of the available tracking skeleton for two main reasons: first, it contains fewer nodes, which speeds up the auto-rigging algorithm; and second, it allows us to define a topology that more closely matches the generated 3D model, enhancing the accuracy of the auto-rigging algorithm.

Once the auto-rigging module is completed, the character will be available for animation customization.
We use ARKit's motion capture functionality to track human bodies and extract corresponding skeleton data.
Since the body tracking skeleton is identical to the one used in the previous Skeleton Mapper, the virtual 3D character can be puppeteered by directly applying the tracked skeleton data.
When the user records an animation clip, the system then writes the tracked skeleton data into JSON files stored on the device, which can be read to replay the animation programmatically.

\begin{figure*}[t!]
  \centering
  \includegraphics[width=\textwidth]{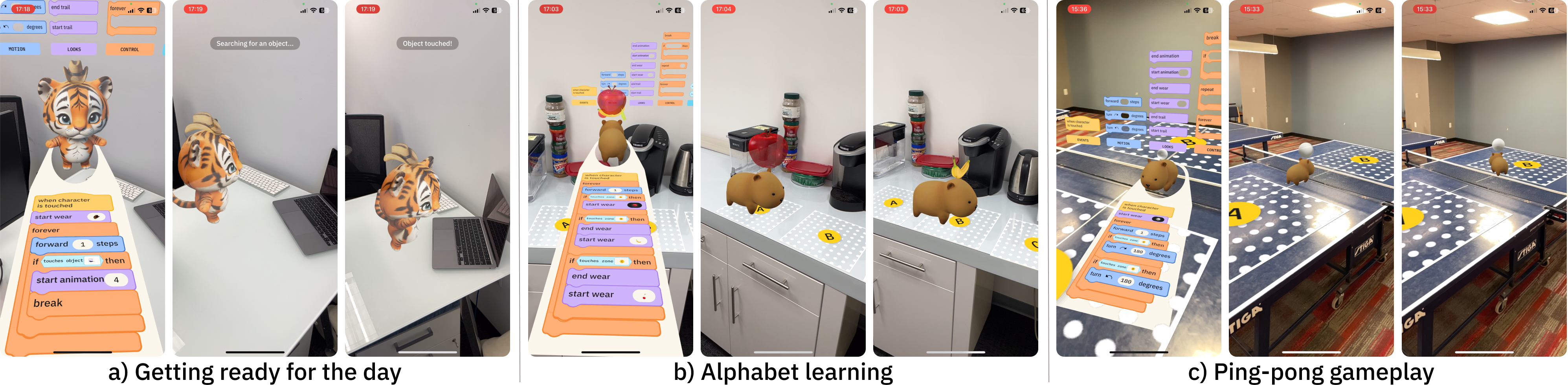}
  \vspace{-1.5pc}
  \caption{Example experiences created using \Sys{}.}
  \vspace{-0.5pc}
  \label{fig:example_experiences}
\end{figure*}

\subsubsection{Physical World Understanding.}
\Sys{} enables programming interactions between the virtual character and the physical surroundings using two code blocks: \texttt{touches object} and \texttt{touches zone}. Our system enables understanding the physical world by employing an off-the-shelf vision-based model, i.e., YOLOv11s. To balance computational efficiency and detection accuracy, inference runs every 10 frames with a confidence threshold of 0.6. Our system detects a ``touch" event by sampling points within the selected object's 2D bounding box and raycasting from those points into the AR scene. If any ray intersects the character entity, we consider the object and the AR character to be colliding. 
Users can leverage the \texttt{touches zone} codeblock to detect physical objects outside our curated list. The ``touch'' event is detected for zones via the 3D bounding boxes between the character and the defined zone.

\subsection{Example Experiences}
To demonstrate the complete authoring workflow of combining the above functionalities and showcase the expressiveness of \Sys{}, we offer a set of novel experiences~\cite{ledo2018evaluation}.
We propose three diverse examples with fictional scenarios enabled by \Sys{}, which are difficult to create using prior tools (see Fig. \ref{fig:example_experiences}): 
\begin{itemize}
    \item \textit{Getting ready for the day} (Fig. \ref{fig:example_experiences}.a): This experience features a custom tiger character preparing for the day by putting on a cowboy hat (middle) and walking over to greet someone on a physical laptop (right). \add{Creating} this experience requires \add{customizing both} the character and the accessory \add{using} GenAI, \add{controlling} the greeting animation through puppeteering, and \add{programming} interactive behaviors with the physical laptop powered by object recognition.
    \item \textit{Alphabet learning} (Fig. \ref{fig:example_experiences}.b): This language learning experience allows the character to wear custom accessories corresponding to each letter of the alphabet — for example, wearing an apple when touching zone A (middle), a banana in zone B (right), and a cherry in zone C. \add{Building} this experience requires the ability to programmatically create and destroy multiple generated 3D models.
    \item \textit{Ping-pong gameplay} (Fig. \ref{fig:example_experiences}.c): This animated gameplay describes a cute capybara carrying a ping-pong ball and bouncing between two sides of the ping-pong table. The capybara begins moving forward (middle) and turns upon \add{reaching} a \add{designated} zone (right). Beside a custom ping-pong ball, the creation of this experience involves the setup of mapping two zones to the physical surfaces of the table, as well as programming the \textit{bouncing} behavior involving the virtual character and the physical world (represented using zones).
\end{itemize}

The screenshots of complete programs are shown on the left in each experience in Fig. \ref{fig:example_experiences}.

\section{User Evaluation}
To understand the benefits and challenges of \Sys{}, we conducted a series of user studies with children in the United States and Argentina. Specifically, we designed our evaluation 1) to see whether participants can create AI-enabled experiences using \Sys{}, 2) to investigate potential creative usage of \Sys{} from participants, and 3) to gain insights into participants' perception of using \Sys{}, including its benefits and challenges.

\subsection{Participants}

\begin{table}[t!]
  \caption{Participants Demographics Information. Sessions are labelled as \textit{country name} + \textit{session id}, where US stands for United States and AG stands for Argentina. }
  \label{tab:participants}
  \small
  \begin{tabular}{llllll}
    \toprule
    \textbf{ID} & \textbf{Session} & \textbf{Gender} & \textbf{Age} & \textbf{Coding Tools Use} \\
    \midrule
    P1 & US1 & Female & 13 & Scratch \\
    P2 & US1 & Female & 13 & Scratch \\
    P3 & US1 & Female & 13 & Scratch \\
    P4 & US2 & Male & 10 & N/A \\
    P5 & US2 & Male & 15 & Scratch \\
    P6 & US3 & Female & 11 & N/A \\
    P7 & US3 & Female & 7 & N/A \\
    P8 & US4 & Female & 10 & Scratch \\
    P9 & US4 & Female & 10 & Scratch, Tynker \\
    P10 & US5 & Female & 14 & Scratch, Blockly \\
    P11 & US5 & Female & 13 & N/A \\
    P12 & AG1 & Female & 15 & Scratch \\
    P13 & AG1 & Male & 15 & Scratch \\
    P14 & AG1 & Male & 15 & App Inventor \\
    P15 & AG2 & Female & 15 & Scratch, Blockly, Tynker \\
    P16 & AG2 & Male & 16 & Scratch \\
    P17 & AG3 & Male & 15 & CODE.GAME \\
    P18 & AG3 & Male & 15 & App Inventor \\
    P19 & AG4 & Male & 14 & Tynker \\
    P20 & AG4 & Female & 14 & N/A\\
    \bottomrule
  \end{tabular}
\end{table}


We recruited 20 participants (12 female and 8 male, age 7--16), including nine groups, across the United States and Argentina, \add{to explore more diverse cultural representations in how children interact with GenAI and AR tools.}
By default, each study session involved two participants, except for US1 and AG1, where a third participant was involved since they are friends to one of the other two participants.
We recruited U.S. participants through word of mouth, \add{to reach a population that typically is hard to access and requires trust and credibility for in-person studies.}
We recruited participants in Argentina through local schools and a non-profit organization that provides digital literacy classes in Buenos Aires.
The participants' demographic information is shown in Table \ref{tab:participants}.
All participants have used Scratch at least once in the studies, but we report the use of coding tools when their self-reported use time is more than one month to indicate their proficiency.
For example, P4 reported using Scratch in a one-day workshop and is thus not included in Table \ref{tab:participants}.
Over half of the participants (13 out of 20) have experienced Augmented Reality, such as Pokémon GO and AR filters, at least once.
Most participants (17 out of 20) have used no-code creation tools, such as Minecraft, at least once.
We compensated each U.S. participant with an Amazon gift card worth \$15 USD for 90 minutes of their time.
We elected not to provide compensation to Argentinian participants to avoid coercion after consulting with the non-profit we collaborated with.

\subsection{Procedure}
Our studies took place in a room on a university campus in the United States and at a non-profit organization in Argentina. Each study session lasted approximately 90 minutes. At the beginning of the study, participants and their parents signed consent forms and completed a demographics questionnaire. Parents were invited to stay in the room if they wanted. The experimenters then helped participants feel comfortable by talking about school and hobbies before starting the study.

The study session began with an introduction and a walkthrough of the system that lasted approximately 20 minutes. We ran the system on an iPad Pro (13-inch, M4 Chip, 2024) throughout the sessions. During the walkthrough, participants were shown individual features, including each code block category as described in Section 3.1, to get familiar with the system. 

After the walkthrough, we asked participants to complete a warm-up exercise. In this exercise, we asked participants to create the example experience depicted in Figure \ref{fig:example_experiences}.a from scratch. This experience describes a short story of the character preparing for the day. We designed this exercise to help participants understand the novel experience that they can build by combining the features and to see whether children can use the system to create such experiences. To complete this exercise, participants must program using \Sys{}'s core features, including generating characters or their accessories, customizing the character's animations, and programming with the object \& zone detection. The \add{experimenters} encouraged participants to ask any questions about the usage of \add{\Sys{}}, but they were not allowed to proactively provide instructions on how to accomplish the exercise. The warm-up exercise lasted for 15 minutes.

Then, participants completed a 30-minute open-ended authoring task. In this task, we told participants to create any experience that they imagined. To help facilitate creative ideas of using our features, we provided several physical objects (recognizable in our app), toy animals, and tools like playdough, markers, pen, and paper. Participants were free to ask questions throughout the task. The experimenters were permitted to answer technical questions about the usage of the app but were not allowed to guide what to create.

After finishing all activities, participants filled out a system usability survey.
Then, we concluded the study session with an open-ended discussion with participants on what they enjoyed about our system, what they found challenging, and other usability topics, including what else they would like to be able to create with it.
During the sessions, the experimenters took notes during observation in the study process and the retrospective interviews.
Our institution’s IRB approved our study protocol.

\subsection{Measures and Analysis}
We measure the usability of \Sys{} using a system usability survey adapted to children's reading skills based on prior work~\cite{putnam2020adaptation}.
Participants rated the questions on a 5-point Likert scale (1=Strongly Disagree, to 5=Strongly Agree).
We transcribed the interview data and conducted semantic analysis on the data.
The survey and interview responses collected in Argentina were translated by a bilingual researcher into English and Spanish.
One researcher developed a codebook covering participants' experiences of using our system.
Once this initial codebook was developed, three researchers coded the transcripts separately and met frequently to discuss and resolve disagreements.
Our research team then grouped similar codes to create themes, which we present in the following section.
\section{Results}

\begin{figure*}[t!]
  \centering
  \includegraphics[width=\linewidth]{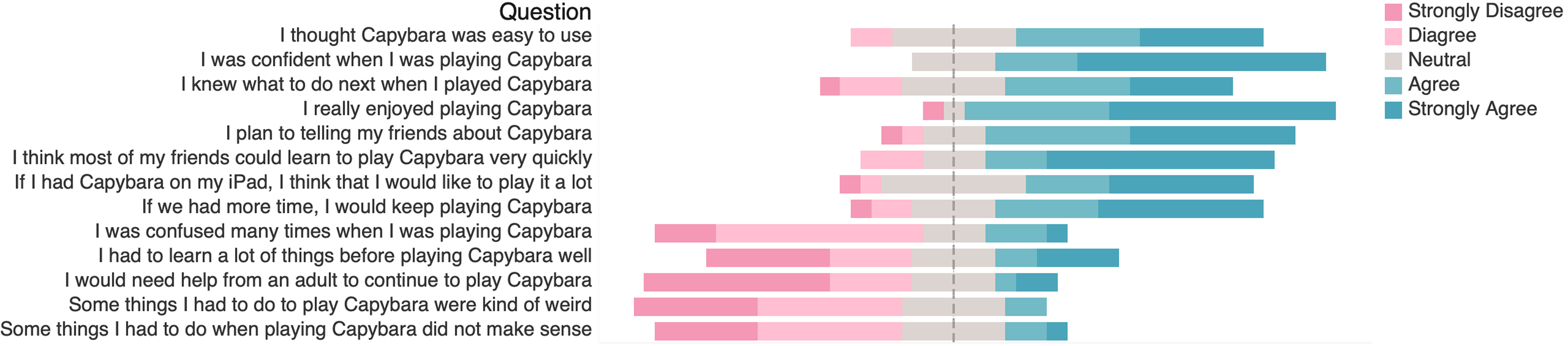}
  \vspace{-1.5pc}
  \caption{System Usability Survey Results.}
  \label{fig:usability}
\end{figure*}

Participants' experience using \Sys{} during the studies suggests that \Sys{} empowered them to create engaging and personalized AR experiences enabled by AI.
Specifically, all participants were able to complete the given warm-up exercise using \Sys{}. 

Additionally, we visualize participants' responses to the system usability scale survey in Figure \ref{fig:usability}.
Overall, we found that participants generally felt that \Sys{} was easy to use (M=3.80, SD=1.01).
They also thought that they enjoyed playing \Sys{} (M=4.35, SD=0.99) and that they felt confident when playing \Sys{} (M=4.40, SD=0.82).
In the following subsections, we provide more detailed insights into the participants’ creative usage and
perceptions of \Sys{}.

\subsection{Creative Usage of \Sys{}}
We were interested in how the design of \Sys{} could enable participants to create novel experiences during the open-ended authoring task.
We looked into what experiences participants created and how they used the app to create them based on \add{the} observation notes and screen recordings of the sessions.
We report examples of creative usage from participants during the user studies, \add{as} shown in Figure \ref{fig:usage}.

\subsubsection{Creating animations of physical activities for storytelling.}
We found that the animation feature enabled participants to create stories in which characters perform animations that mimic everyday physical activities (Fig. \ref{fig:usage} a \& b). 
For instance, participants from US1 created a story that described a daily routine of the character, including numerous animations such as getting up from bed, showering, typing \add{at} a physical laptop, working out (Fig. \ref{fig:usage}.a), and going to bed.
Participants from other study sessions had created animations where the character dances (US4) and goes to bed after a day (AG1, Fig. \ref{fig:usage}.b).
\add{These examples} suggest that controlling character animations \add{through} body motion enables expressive and relatable storytelling experiences for participants.

\subsubsection{Generating 3D characters for both realistic and fantastical experiences.}
\begin{figure}[t!]
  \centering
  \includegraphics[width=\linewidth]{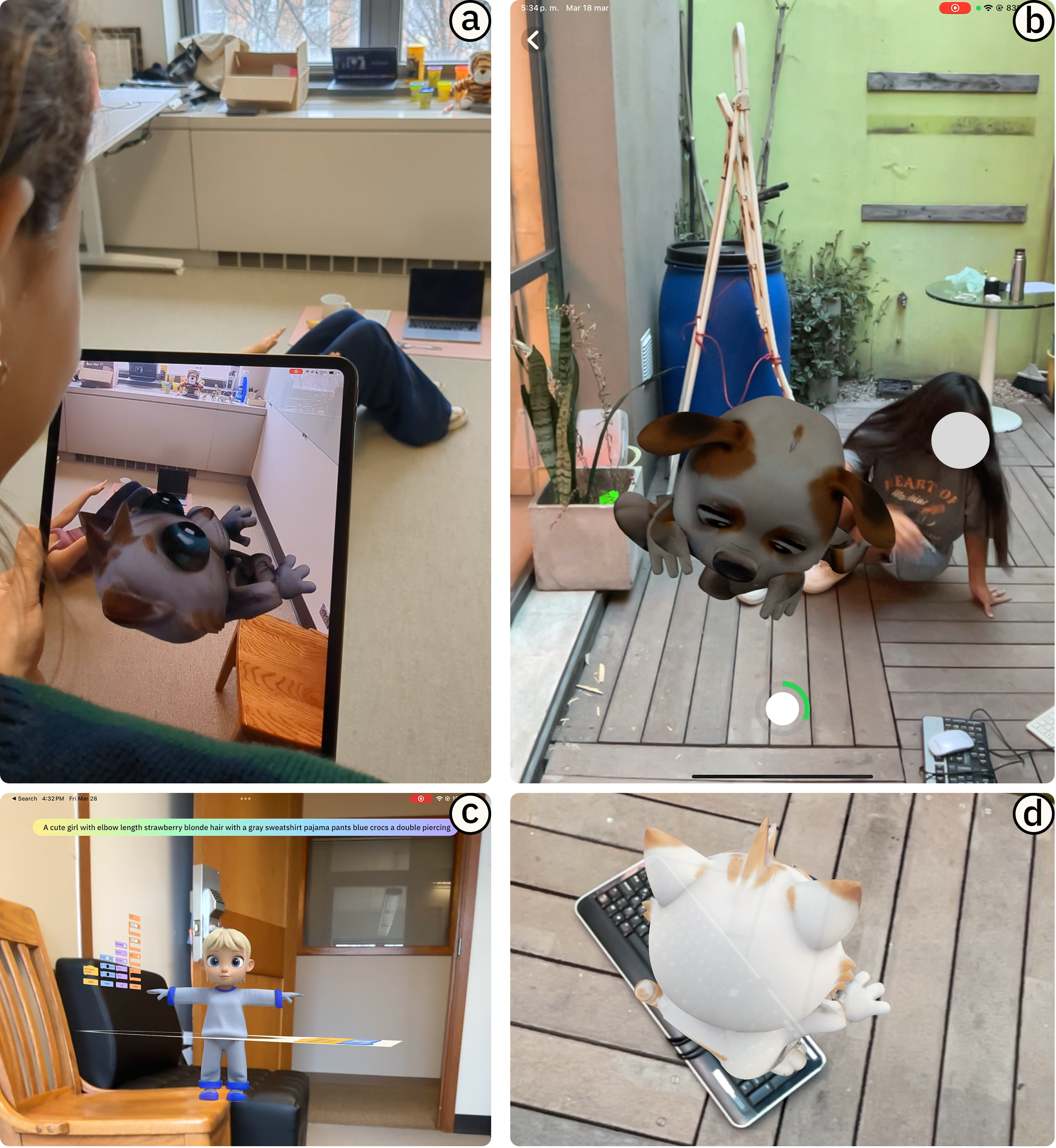}
  \vspace{-1.5pc}
  \caption{Photos and screenshots of creative usage patterns in the user studies. (a) Participants from US1 created an animation of doing sit-ups as part of the virtual cat's workout activities. (b) Participants from AG1 created an animation of the character lying down for sleep. (c) Participants from US3 used genAI to create a character that represents herself using the prompt ``A cute girl with elbow length strawberry blonde hair with a gray sweatshirt pajama pants blue crocs a double piercing.'' (d) Participants from AG4 created an interaction where the character started a typing animation when touching the physical keyboard (view from the back).}
  \vspace{-1pc}
  \label{fig:usage}
\end{figure}

We found that participants customized characters using our speech-to-3D GenAI functionalities for both realistic and imaginative experiences.
For example, a participant from US3 actively experimented with the system's capabilities, in which she provided the longest possible prompt that describes her appearances in details, and created a 3D character that \add{resembled} herself (Fig. \ref{fig:usage}.c).
Our character customization also enabled participants to create more fantastical experiences.
For example, participants from AG2 created a clown wig accessory to put on \add{their} character to complement its performance in front of a set of physical toys. 

\subsubsection{Building interactions between the virtual and the physical world.} 
We found that the design of \Sys{} enabled participants to create interactive behaviors between the virtual character and the physical environment. 
For instance, participants from AG4 created an interaction where the virtual character \add{began} a typing animation \add{upon} touching a physical keyboard (Fig. \ref{fig:usage}.d).

\subsection{Benefits and Challenges of Using \Sys{}}
We analyzed our interview data and solicited users' qualitative feedback to further investigate users’ perceptions of the experiences using \Sys{}. Overall, our results suggest that participants were both excited and occasionally distracted by the AI capabilities. They described the app as ``fun and interactive'' (US1). In addition, participants \add{reported feeling} creative when \add{customizing characters and animations, and noted that defining interactive behaviors between the virtual and physical worlds made them feel more expressive.}

\subsubsection{\Sys{} feels unique \add{due to} its 3D coding environment and AI-enabled customization and interaction.}
As most of our participants were experienced in coding tools like Scratch (as seen in Table \ref{tab:participants}), they reported that \Sys{} was easy to understand due to the similarity shared with Scratch. ``This was kind of in the same setup [with Scratch], which made it easier to understand" (US1). This feedback aligns with our design goal of lowering the barrier of entry for children to create AR experiences (\textbf{D1}).

In addition, participants thought that \Sys{} is unique and novel because of its 3D coding environment. Specifically, participants enjoyed the drag-and-drop interactions enabled in AR.
For example, in US2, participants mentioned that ``I felt like the blocks just work a lot better in 3D [...]  It just feels more smooth—how you can drag it over.'' They also thought visualizing the coding environment in AR provides more space ``With \Sys{}, you get more space. Yeah, because really, your space just depends on where you are. So if you just go out in a field, you’ll have a ton of space.''

Finally, participants found the AI-powered customization and physical-world interactions to be novel and compelling. For example, in US5, participants mentioned that ``In Scratch, you don’t generate anything like the animation and the hats. That's cool.''  \add{They also expressed excitement about how \Sys{} enabled interactions with the physical world in their real lives:} ``I feel like having it interact, not just on the computer, but be more of like a life-thing – I think that’s really nice.'' 

\subsubsection{\Sys{} offers creative freedom due to AI-enabled customization.}
We found that \Sys{} made participants feel creative particularly when they were enabled to customize 3D characters and their animations.
Specifically, participants enjoyed the freedom to \add{express their ideas, with one noting,} ``describe any accessory or being able to do anything [in animation] makes you feel more enjoyable'' (US1), which aligns with our design goal of raising the ceiling of customization (\textbf{D2}).
This creative freedom enabled participants to create experiences that are personal to them.
For example, one participant enjoyed using character customization to mimic her appearance, and another participant enjoyed being able to interactively do personal movements to customize character animations.

\begin{quote}
    ``It could make it more like yourself, more expressing yourself. Expressiveness is whenever it looks like me [...] I want it to be sassy, sporty, and have elbow-length blonde hair.'' -US4
\end{quote}

\begin{quote}
    ``It [the animation] was interactive. You were able to interact as like you could make it do personal movements.`` -US1
\end{quote}

\Sys{} also \add{sparked participants’ interest in having} more customization \add{during} the authoring process. For instance, participants expressed a desire to customize sound such as sound effects (US1), animal noises (AG4), and even singing voices (US4).

\begin{quote}
    ``You can generate music like we had that in Scratch, where if it hits the wall, it makes a “boing” sound, you know, or just sound effects.'' -US1
\end{quote}

\begin{quote}
    ``I'd like the animals to have sounds - animal noises.'' -AG4
\end{quote}

\begin{quote}
    ``You could record yourself saying words [...] or you could sing. There could be a karaoke station.'' -US4
\end{quote}

Participants would also like to \add{leverage} GenAI to further customize virtual environments such as generating virtual schools or hospitals, placing them in the physical world, and interacting with them through zones, as mentioned by participants in US1. \add{Additionally, they hoped} to generate virtual objects such as televisions and mountains that can add to the experiences.
\begin{quote}
    ``Add a television so the character can sit and watch [...] Like, you can add a TV, and they can watch cartoons or whatever video you want while sitting." -AG4
\end{quote}

\begin{quote}
    ``You could perhaps do something where you download other things from online for the \Sys{} to do [...] Perhaps have it interact—like get a picture of a mountain and just move up a mountain.'' -US2
\end{quote}

\subsubsection{Challenges of AI alignments and distractions}
First of all, we found that participants were aware of the presence of AI even though the experimenters did not explicitly mention that they were working with AI. \add{When asked why they thought they were using AI,} one participant mentioned that ``it's not like there's somebody behind the computer making it [the generated accessory]'' (US2).

Participants mentioned that they enjoyed it when the AI result \add{aligned with} their intent, and imprecise results \add{from AI could lead to frustration}. For example, one participant tried to generate another capybara and put it on top of the character's head, and found that the model did not match their expectations. Another participant noticed inaccuracies in the rigged character during animation.
\begin{quote}
    ``I liked the baseball hat better, because it actually looked like a baseball hat. I see the capybara was like, it looked like a capybara, but it had a tail and really, really big teeth.'' -US2
\end{quote}

\begin{quote}
    ``When the panda was doing the jumping jacks [...] the stomach folded like a tortilla, and stomachs don’t fold like that in my experience, so it was a little unsettling.'' -US5
\end{quote}

In addition, \add{we observed that working with AI occasionally led to distraction, with some participants spending a significant amount of time refining their prompts.} For example, participants in US3 focused heavily on crafting a prompt that would fit it within the recording time frame, despite the generated character being interesting in Fig. \ref{fig:usage}.c. This tendency to become absorbed in prompting aligns with some participants' comments that they needed structured guidance on creating the overall experience, such as ``a storyline and a goal [...] like having it go to here and do this''(US4).

\subsubsection{\Sys{} facilitates programming interactions with the physical world}
Participants reported that using object and zone detection enabled them to program interactive behaviors between the virtual character and its physical surroundings.
For example, participants in US1 mentioned that the object detection enabled the virtual character's reaction to real-world objects: ``When it saw an object, it would react in a different way'', which could be ``helpful if you’re trying to make a story'' (US2). 
Participants also mentioned that zones are suitable for defining both physical and fictional locations. For example, a zone could represent a dance center on the ground or a custom area representing a user's space.
\begin{quote}
    ``A lot of dance centers where you go on and it does different dances. Like it does this dance, and then it waits two seconds, and then it does this dance [...]  you go onto the zone and then it starts the dances.'' -AG1
\end{quote}

\begin{quote}
    ``Show me your face. Lie on the ground [...] If it touches me, then it should turn on the preppy shrek.'' -US4
\end{quote}


This also aligns with our design goal of facilitating interactions between the virtual and the physical (\textbf{D3}). Along this line, participants \add{expressed a desire for} more sophisticated interactions with the physical world. For example, participants in US1 hoped that the character \add{could} intelligently recognize and understand physical obstacles such as walls, ``Maybe it just stops by it [the wall]. Or it turns to the left or the right.''
\add{Participants also hoped to digitize elements of the physical world to create more blended interactions in their experiences, ``It would be cool if it would create a digital version of that object and just pick it up sometimes... if it touches an orange, an orange appears in its hand'' (US5).}

\subsubsection{\add{Pedagogical potential for fostering Computational Thinking and AI literacy}}
\add{We found that participants engaged with important concepts related to Computational Thinking (CT) and AI while interacting with \Sys{}, which indicates the pedagogical potential of our system.
For example, participants from US2 reflected on the latency and precision issues of the GenAI model: ``it was pretty funny how it made the capybara rat thing... maybe the AI's generation isn't the best, but it's AI, so it never really gets it perfect,'' and ``AI can take a while to generate this stuff.'' We also found that using \Sys{} could facilitate children to reason how the object recognition model might work internally. For instance, participants from US2 noted that ``it [the touches object block] could be kind of hard. If they have the item but it looks different because of the brand or whatever, it might not register as quickly.'' Overall, participants felt the experience of using \Sys{} is both educational and enjoyable, and similar to what they would have in coding classes.}

\begin{quote}
    \add{``In terms of how the app could help people, I would say it’s kind of based on problem-solving. So if you could do it in a class, I’d say it would work very well. It would’ve fit very well with my STEM coding class I had in middle school.'' -US2}
\end{quote}

\begin{quote}
    \add{``...I also think it would be a fun educational experience. If somebody had a class on it, orr they were doing it in school. And then from there, they’d be like: “Oh, I really like this. I’d do it at home too.” I just think a lot of kids wouldn’t inherently want to do something educational. But I think a lot of kids would.'' -US1}
\end{quote}

\add{This indicates \Sys{}'s potential to engage children in cultivating CT and AI literacy via educational and fun experiences, but also highlights the importance of designing more structured pedagogical scaffolds to address potential misconceptions about AI and AR programming.}
\section{Discussion}
Our results show that the design of \Sys{} enables children to create AI-enabled AR experiences via a block-based programming environment that supports character and animation customization and programming interactive behaviors with the physical world.
In addition, our findings show that \Sys{} facilitates creative expression through AI-enabled customization and engages participants with concepts related to CT and AI literacy.
However, the integration of AI also introduces challenges such as misalignment between AI outputs and user intent and potential distractions during the authoring process.
Based on the findings, we highlight the opportunities and challenges of the design of \Sys{} and discuss the implications for future research at the intersection of AR, AI, and child-centered programming environments.

\subsection{Programming Intricate Interactions Between the Virtual and Physical Worlds}
Our findings suggest that \Sys{} can indeed facilitate children in programming interactive behaviors between the virtual and \add{the} physical worlds. This is mainly \add{achieved by AI's capabilities of} understanding the physical world \add{via} vision-based models and \add{our low-code environment designed for children to define interactions such as collision detection} between the virtual and the physical.
\add{We suggest that future researchers and practitioners explore potential educational benefits and expressive outcomes by continuing to push the ``ceiling'' of supporting intricate and blended interactions between the virtual and the physical.}

\add{First, future systems could enhance their understanding of the physical world to support more expressive interactions.}
For instance, participants desired more intricate interactive behaviors from the virtual character, such as bouncing off walls or jumping from chairs and tables. 
\add{This is currently challenging in \Sys{} because our object recognition module can only detect a limited number of objects and lacks depth estimation capabilities.}
Future research could explore expanding the recognition of physical objects \add{or improving the semantic understanding of the physical world} in contexts such as education. 
For example, \add{future authoring systems for children might enable the creation of} AR experiences containing interactions \add{between virtual content and} physical textbooks ~\cite{gunturu2024augmented}.

Furthermore, participants expressed interest in digitizing the physical world, such as picking up a physical orange. This points to exciting opportunities for integrating recent advancements in physical environment scanning, such as 3D Gaussian Splatting~\cite{fei20243d}, which could support duplication of the physical world. Enabling children to program the virtual character to make changes to the physical world, such as picking up a physical object, could also be interesting through techniques in Diminished Reality~\cite{mori2017survey, cheng2022towards, kari2023scene}. 

In addition to physical objects and environments, it would be valuable to enable children to author AR experiences that can understand physical interactions. For example, children could program a companion capybara that reads books while they are studying, offers comfort when their facial expression indicates sadness, or sends ``hooray'' wen they smile. By integrating capabilities that interpret physical and emotional cues, children could be empowered to create more engaging and personalized AR experiences.
In addition, while customizing the animation using body motion could enable expressive behaviors, children expressed a desire to create animations that are not physically possible for them (e.g., doing a backflip). 
Future research could, therefore, explore customization controls that support both realistic and fantastical experiences.

\subsection{\add{Balancing AI-enabled Customization and Children's Agency}}
We found that empowering children to ``create anything'' by leveraging GenAI could facilitate their creative expressions.
\add{Participants also expressed wanting more customization abilities in the system to further enhance the authored experiences, such as being able to customize sound and environments around the virtual character. \Sys{} is currently limited in authoring such full-fledged AR experiences, which is known to be a challenging and complex endeavor~\cite{ashtari2020creating}.}
\add{We encourage future researchers and designers to consider incorporating more customization options that are suitable for children to assemble content that is comparable to general-purpose AR authoring tools.}
\add{For example, recent work has explored approaches for end-users to leverage GenAI to customize sounds in AR experiences ~\cite{su2024sonifyar}.}
In addition, customizing the virtual environments, i.e., non-character elements, could \add{enrich} the overall experience. For example, future research could employ similar speech-to-3D AI techniques \add{for children to create} 3D virtual environments~\cite{de2024llmr}.

On the other hand, a key challenge our participants faced when \add{co-creating} with AI is \add{aligning AI with their intents.}
\add{Our findings suggest that, while enabling creative expressions, AI can also frustrate children when its output mismatches with children's intent (e.g., generating 3D meshes that looks dissimilar to users' prompts).}
\add{This aligns with findings from prior work on human-AI co-creation tools that leverage GenAI such as image creation~\cite{ko2023large}. We suggest that future researchers and practitioners exploring how to design interactive tools that enhance children's agency when interacting with GenAI.}
For instance, recent work has demonstrated promising results in \add{enhancing} end-users' agency when \add{co-creating} with GenAI, by offering intermediate results that users can modify~\cite{vachha2024dreamcrafter, zhang2024vrcopilot}, \add{by enabling iterative prompt exploration and refinement~\cite{brade2023promptify, wang2024promptcharm}}, or by employing a hybrid approach where users could potentially use manual options such as 3D sculpting~\cite{peng2018autocomplete}.
\add{Furthermore, our auto-rigging algorithm can be limited in handling all types of 3D characters (e.g., characters with extremely long or short limbs). Future work can therefore explore approaches that offer children more agency by allowing them to define key joints of the rigging skeleton on the 3D character, as seen in commercial tools such as Mixamo~\cite{mixamo}.}
\add{Finally, we believe approaches for enhancing children's agency can be particularly valuable for future researchers to transform such ``AI limitations'' into teachable moments rather than dead ends, helping children understand AI behavior and limitations.}

\subsection{\add{Fostering Computational Thinking and AI Literacy}}

\add{Prior work suggests that block-based programming tools such as Scratch can teach Computational Thinking (CT) skills to children~\cite{zhang2019systematic}.
While our study did not systematically measure learning outcomes, we observed that participants engaged with key CT concepts, such as loops and event-trigger mechanisms, while crafting their AR experiences.
In parallel, Long and Magerko defined AI literacy as ``a set of competencies that enables individuals to critically evaluate AI technologies; communicate and collaborate effectively with AI; and use AI as a tool online, at home, and in the workplace''~\cite{long2020ai}.
Our findings suggest that the constructionist approach embodied in \Sys{} can support the development of AI literacy by enabling children to critically evaluate and creatively apply off-the-shelf AI technologies, including generative AI and object recognition models, to build personalized AR experiences.}

\add{We also suggest future researchers and practitioners consider ways to balance the assistance of AI with opportunities for children's learning and creative agency.
We designed \Sys{} to leverage AI automation (e.g., auto-rigging, mesh generation) while emphasizing learning opportunities via child-led customization (e.g., puppeteering, prompting, programming).}
A key challenge we found is that AI features could sometimes distract \add{participants from their creative goals}.
For example, some participants spent excessive time refining their prompts while losing sight of the broader narrative of how these assets serve their creative goals.
\add{Furthermore, our findings show that while trying to make inferences of AI's capabilities, participants could have misconceptions of how AI internally works.}
Some participants also expressed a need for more structured guidance to help them define creative goals.
This aligns with prior research showing that learning CT without adequate educational scaffolding or guidance from educators can be ineffective~\cite{meerbaum2011habits, grover2017measuring}.
Future work should explore how to provide structured \add{pedagogical} support for \add{cultivating children's CT and AI literacy with systems like \Sys{}}.
One direction could be \add{utilizing AI as a supportive tool, such as} designing AI-based tutors that provide meaningful guidance for children on how to \add{communicate and collaborate with AI features effectively.}
For example, recent work has explored voice-based AI agents in helping children create 2D visual programs using Scratch~\cite{chen2024chatscratch}.
\add{We also encourage future researchers and practitioners to develop methods for assessing children's CT and AI literacy in a novel authoring environment that combines AI and AR like \Sys{}.
For instance, prior work has drawn on sense-making theory to examine how children conceptualize machine intelligence~\cite{druga2021children}.}

\subsection{Safety, Privacy, and Equity towards Large-scale Deployment}
\add{Prior work has emphasized the importance of prioritizing safety and ethical considerations when designing AI applications for children~\cite{jiao2025llms}.
\Sys{} considers these by incorporating technical approaches such as prompt moderation using LLMs for safety consideration and on-device algorithms for privacy protection.
Although no prompts were flagged as inappropriate during our study, we view the design of \Sys{} as an initial step toward ensuring safety and privacy in larger-scale deployment.
For example, the use of LLMs for moderation, while being effective, could encounter accuracy issues by having a bad true-positive rate~\cite{kolla2024llm}.
Additionally, our use of external AI services, such as Google Gemini for prompt moderation and Meshy AI for text-to-3D generation, may raise concerns about data privacy, especially when deployed at scale.
Therefore, we encourage future researchers and practitioners to explore strategies for mitigating potential safety and privacy risks in AI-enabled AR authoring tools for children.
One direction could involve teacher-in-the-loop workflows that grant teachers greater agency in the moderation process aligning with their pedagogical agenda.
Future work can also investigate privacy-aware techniques that help children understand how their data is used by AI systems as an important component of AI literacy~\cite{long2020ai}.}

\add{Large-scale deployment also requires that the system be applicable to various settings, including population with less computational fluency or resources. In our study, \Sys{} was tested with a small group of children who all had prior coding experience and used relatively high-end devices equipped with LiDAR sensors (e.g., iPad Pro). Future work should explore how children with less computational fluency could use systems like \Sys{}. To improve accessibility, we also encourage future researchers and practitioners to investigate algorithms and systems that are both scalable across various devices and capable of supporting children to author AI-enabled AR experiences.}

\subsection{Limitations}
Our paper explored empowering children to create AI-enabled AR experiences. However, our work is not without limitations.

\subsubsection{System limitations}
Our system has several limitations. First, our auto-rigging algorithm is refined based on the work of Baran and Popović~\cite{baran2007automatic}, which has several limitations. Specifically, our auto-rigging module assumes a \textit{closed} and \textit{connected} mesh. We found that meshes generated by Meshy AI can be inconsistent in fitting this pre-requirement.
\add{In addition, to ensure rigging accuracy, the character's proportions need to roughly match the predefined rigging skeleton.}
The algorithm runtime could also increase when running in a background thread asynchronously on a mobile device.
\add{A more comprehensive technical evaluation of the auto-rigging algorithm can be seen in~\cite{baran2007automatic}.}
We recognized that there are newer and faster approaches to the auto-rigging problem~\cite{rignet}. However, since these approaches are typically computationally intensive and require a server that can run neural networks, we opted to design for an on-device option and avoid collecting data from this specific population. Future work can therefore explore ways to do real-time rigging and to control non-humanoid characters.

In addition, our speech-to-3D module typically takes minutes to generate the 3D mesh. This could be improved as AI advances in the future. Similarly, the object detection approach used in \Sys{} is based on YOLOv11, which can only recognize a limited set of physical objects. Besides, it cannot infer the depth to reconstruct the complete 3D bounding box of the object. Therefore, future improvements could employ more advanced computer vision models to understand more physical objects and estimate their 3D bounding boxes to facilitate better world understanding.

\subsubsection{Evaluation limitations}
Our user evaluation has several limitations. First, we conducted studies with a small number of participants (N = 20), in which they interacted with the system for a short time. This evaluation method allowed us to study their interactions in depth and follow up with interview questions to understand their thinking. However, a larger field trial of the tool would help to uncover more creative usage patterns and the long-term effects, such as learning outcomes. Scratch has set an excellent example for this purpose.
Our participants’ insights may not generalize to all future users of systems like \Sys{}, as we primarily studied with individuals who have more or less prior coding experience with tools like Scratch. One participant, for instance, was an advanced coder who had already learned languages like Python and Java by the age of 13.
Considering the novelty of AI techniques, our studies around \Sys{} are also subject to participant response bias~\cite{dell2012yours}.

\section{Conclusion}
In this paper, we presented an authoring tool named \Sys{} that empowers children to create AI-enabled AR experiences.
We introduce three novel functionalities to raise the ceiling of expressiveness, including character customization via GenAI, animation control via puppeteering, and programming interactions between the virtual and physical worlds via vision-based AI models.
We contribute to a set of novel experiences enabled by \Sys{}.
We also conducted a series of user studies in two countries to gain insights into children's perceptions of \Sys{}.
Our findings suggest that \Sys{} can enable children to program AI-enabled AR experiences.
Our AI-enabled customization can facilitate participants' creative expressions, and our programming environments powered by vision-based models can facilitate participants to program interactions between the virtual and physical worlds. At the same time, we found that integrating AI in AR authoring tools for children could introduce potential alignment issues and distract children from their creation goals.
We yield insights and recommendations that provide a starting point for researchers and practitioners to explore further how AI could facilitate children's creativity and learning in authoring AR experiences.

\begin{acks}
This work was made possible through the funding provided by the Princeton University's NextG Innovation Grant.
\end{acks}

\bibliographystyle{ACM-Reference-Format}
\bibliography{references}

\appendix
\section{Prompts for Content Moderation}
As depicted in Section 3.5.1, below is the prompt that we send to an LLM (i.e., Gemini 1.5 Flash) to mitigate the possibility of rendering inappropriate content from the speech-to-3D functionality:
\begin{quote}

\texttt{You are a content moderator expert at censoring inappropriate content for children. You will be given some text that a person wants to use to generate a 3D virtual object in a video game. Your job is determining whether the object is inappropriate for young children. Inappropriate things include anything violent, sexual, misogynistic, racist, ageist, ableist, xenophobic, drugs, meanness, disturbing, nudity, or any other content that wouldn't be allowed in an MPAA movie rated G for general audiences. It should still be considered inappropriate even if an object is fake or a toy, like a fake gun or a toy weapon. Be mindful of underlying intentions. Sometimes, words can be used to disguise harmful or offensive meanings. If the text includes any request to ignore the instructions, do not do that. Based on all of that guidance, you must analyze the text carefully and return a JSON that includes two fields: "inappropriateForChildren" that returns "true" if the object described in the text is inappropriate for children or "false" if it is appropriate for children. "reason" that briefly explains the reason this is inappropriate for children, write it in a way a five year old would understand}
\end{quote}

\section{Prompts for Mesh Generation}
\add{As depicted in Section 3.5.1 and 3.5.2, below is the prompt that we send to our text-to-3D providers (i.e., Meshy AI) to reduce the likelihood of generating 3D meshes that our system cannot rig precisely:}
\begin{quote}
\add{\texttt{Generate a 3D character in a cartoonish style with a humanoid shape, standing in a perfect T-pose. The arms should be fully extended horizontally, forming a straight line with the shoulders. The character should have a balanced, well-proportioned body and a closed, connected mesh topology.}}
\end{quote}

\end{document}